\documentclass[journal=nalefd,manuscript=letter,nofootinbib]{achemso}

\usepackage[T1]{fontenc}
\usepackage[latin1]{inputenc}
\usepackage{lmodern}
\usepackage{amsmath}
\usepackage{amssymb}
\usepackage{pst-all}
\usepackage{psfrag}
\usepackage{graphicx}
\usepackage{dcolumn}
\usepackage{bm}
\usepackage{epsfig}

\title{Universal Distance-Scaling of Non-radiative Energy Transfer to Graphene}

\author{L.~Gaudreau}
\affiliation[ICFO]{ICFO - The Institute of Photonic Sciences, Mediterranean Technology Park, Av. Carl Friedrich Gauss 3, 08860 Castelldefels (Barcelona), Spain}
\author{K.~J.~Tielrooij}
\affiliation[ICFO]{ICFO - The Institute of Photonic Sciences, Mediterranean Technology Park, Av. Carl Friedrich Gauss 3, 08860 Castelldefels (Barcelona), Spain}
\author{G.~E.~D.~K.~Prawiroatmodjo}
\affiliation[ICFO]{ICFO - The Institute of Photonic Sciences, Mediterranean Technology Park, Av. Carl Friedrich Gauss 3, 08860 Castelldefels (Barcelona), Spain}
\author{J.~Osmond}
\affiliation[ICFO]{ICFO - The Institute of Photonic Sciences, Mediterranean Technology Park, Av. Carl Friedrich Gauss 3, 08860 Castelldefels (Barcelona), Spain}
\author{F.~J.~Garc\'{\i}a~de~Abajo}
\affiliation[CSIC]{IQFR - CSIC, Serrano 119, 28006 Madrid, Spain}
\author{F.~H.~L.~Koppens}
\affiliation[ICFO]{ICFO - The Institute of Photonic Sciences, Mediterranean Technology Park, Av. Carl Friedrich Gauss 3, 08860 Castelldefels (Barcelona), Spain}
\email{frank.koppens@icfo.eu}

\begin{document}

\begin{abstract}
The near-field interaction between fluorescent emitters and graphene exhibits rich physics associated with  local dipole-induced electromagnetic fields that are strongly enhanced due to the unique properties of graphene. Here, we measure  emitter lifetimes as a function of emitter-graphene distance $d$, and find agreement with a universal scaling law, governed by the  fine-structure constant.  The observed energy transfer-rate is in agreement with a $1/d^4$ dependence that is characteristic of 2D lossy media. 
The emitter decay rate is enhanced 90 times (transfer efficiency of $\approx 99\%$) with respect to the decay in vacuum at distances $d\approx 5$ nm. This high energy-transfer rate is mainly due to the two-dimensionality and gapless character of the monoatomic carbon layer. Graphene is thus shown to be an extraordinary energy sink, holding great potential for photodetection, energy harvesting, and nanophotonics.
\end{abstract}

Graphene, a genuinely two dimensional material composed of a single layer of carbon atoms, has rapidly generated great interest since its experimental isolation \cite{Novoselov:2004p2258} thanks to its extraordinary electronic, optical, and mechanical properties. Due to its gapless band structure and linear electronic energy dispersion \cite{CastroNeto:2009p2250}, graphene exhibits frequency-independant light absorption over a broad spectral region in the visible and infrared \cite{Mak:2008p3352}, which is governed only by fundamental material-independent constants: its absorbance is given by $\pi\alpha\approx2.3\%$, where $\alpha\approx1/137$ is the fine-structure constant \cite{Nair:2008p1472}. This new material displays high room-temperature mobilities \cite{Mayorov:2011di}
  up to 250,000\,cm$^{2}$V$^{-1}$s$^{-1}$, electrically tunable carrier concentration, and bipolar field response \cite{Novoselov:2004p2258}. These optical and electronic properties have promoted the use of graphene for a multitude of opto-electronic applications \cite{Bonaccorso:2010p1871} and as a potential platform to exploit light-matter interactions under ambient conditions \cite{Koppens:2011p3370}.

Recently, efficient energy-transfer from bio-molecules to graphene has  sparked tremendous interest for sensing purposes \cite{Liu:2010gw,Dong:2010cu,Chang:2010gf,Zhang:2011if,Chen:2011gg,Liu:2011jp,Wang:2011da}. From a fundamental point of view, the near-field interaction between an emitter and a purely two-dimensional material is particularly interesting because it allows for the exploration of new limits of light-matter interactions \cite{Koppens:2011p3370,Nikitin:2011ey}. First of all, due to the two-dimensional and gapless character of graphene the magnitude of the non-radiative coupling is strongly enhanced relative to other lossy materials. This strong coupling has been predicted to produce substantial energy transfer up to distances of 30\,nm \cite{Swathi:2008p1605}. Additionally, the non-radiative energy transfer rate $\Gamma_{\rm nr}$ as a function of distance $d$ between the emitter and graphene has been predicted\cite{GomezSantos:2011p1506,Swathi:2008p1605} to scale as $d^{-4}$, in contrast to the energy transfer of emitters near conventional semi-infinite materials, where the magnitude scales with $d^{-3}$. Finally, the emitter decay rate enhancement follows a universal scaling law governed by the fine-structure constant and the ratio of $d$ and the emitter wavelength $\lambda$ \cite{GomezSantos:2011p1506}.

From an application point of view, we envision novel types of hybrid systems that combine graphene with strong light absorbers/emitters (e.g., quantum dots and fluorescent molecules) to enhance the intrinsically weak graphene absorption, and thereby, also the device efficiency for light harvesting and photodetection.  Finally, the scaling of the decay rate as a function of distance makes the emitter-graphene system promising as a nanoscale ruler, as proposed in Ref.\ 17.
 Further advances towards these new effects and applications require a quantitative experimental study, which has not yet been realized.


\begin{figure}
\begin{center}
\includegraphics*[scale=0.8]{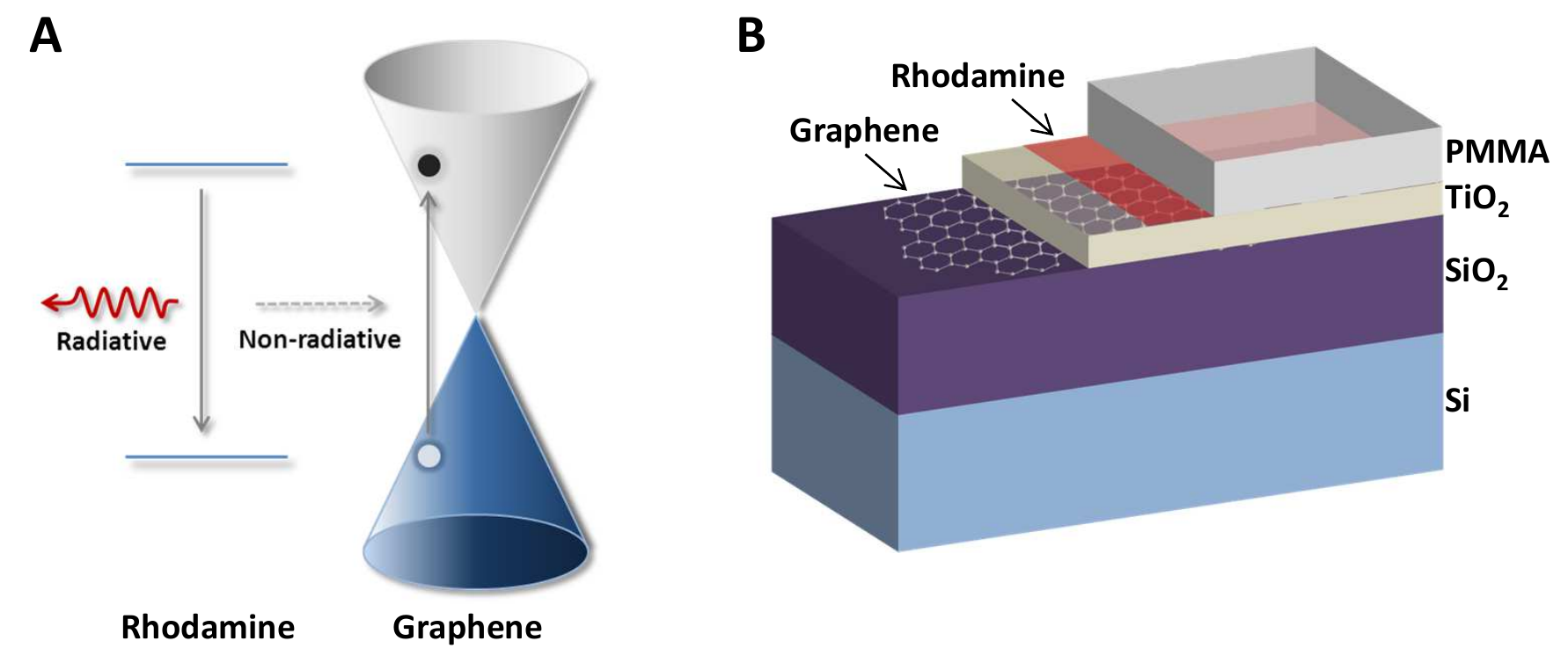}
\caption{(A) Energy diagrams of an optical emitter and a nearby graphene monolayer sheet. Upon optical excitation of the emitter, the relaxation of the excited state occurs for short distances primarily through non-radiative decay by dipolar coupling to electron-hole pair transitions in the carbon layer, and to a lower extent through the emission of radiation. This leads to emission quenching and a shorter lifetime of the emitters. (B) Schematic representation of the sample structure. The separation between the emitters and the graphene flake is implemented through iterative atomic layer deposition of a TiO$_2$ spacer layer, as explained in the text.}
\label{fig:1}
\end{center}
\end{figure}

The efficient energy transfer between light absorbers/emitters and graphene relies on the strength of their near-field interaction. Recent theoretical studies \cite{GomezSantos:2011p1506, Swathi:2008p1605, Velizhanin:2011p1796} suggest that this interaction is mediated by non-radiative coupling between the emitter dipole and electron-hole pair excitations in graphene (i.e., F\"{o}rster-like energy transfer, see Figure \ref{fig:1}A), which in turn results in higher decay rates and emission quenching (i.e., the energy released from the emitter, that would otherwise be re-emitted as light, is absorbed by the graphene). This picture is consistent with recent experiments that have demonstrated that the fluorescence of nitrogen vacancy centers in diamond \cite{Stohr2012}, CdSe/ZnS nanocrystal emitters \cite{Chen:2010p1464}, fluorescent dyes \cite{Treossi:2009gy,Kim:2010p3299}, and dye-labeled DNA \cite{Huang2012} is quenched by the presence of graphene.

In this Letter, we perform a quantitative study of the near-field interaction between graphene and nearby emitters. Specifically, we measure the decay rate of a thin layer of rhodamine fluorescent molecules (emitters) coupled to a monolayer graphene flake as a function of emitter--graphene distance. We show that graphene induces a dramatic change in emitter lifetime, increasing the decay rate of the excited molecules by up to 90 times. This corresponds to $>$99\% of the energy stored in the emitters being transferred to graphene. We compare the results with a simple (but rigorous) analytical model and find good agreement without any fitting parameters. The observations reveal that the strong near-field interaction originates from the unique properties of graphene: its gapless and two-dimensional character as well as its charge carriers being relativistic massless Dirac fermions. As a result, the non-radiative energy transfer to graphene is governed by the fine-structure constant and exhibits a $1/d^{-4}$ universal scaling with the distance $d$ between emitters and graphene.

\section*{Experiment}
\label{sec:Experiment}

We measure the optical-excitation lifetime of a thin layer of rhodamine molecules ($<1$\,nm thick \cite{Heinz:1982p1485}) by probing spatially and temporally resolved fluorescence using a home-built confocal microscope. A pulsed green laser ($532\,$nm wavelength) at 40 MHz excites the molecules and their fluorescence is recorded with an avalanche photodiode detector (APD). Through time-correlated photon counting with a Picoharp time correlator, we obtain the lifetime of the emitters. The sample consists of a monolayer graphene flake placed on top of a Si-SiO$_2$ (285 nm) substrate. The graphene is covered by a TiO$_2$ spacer layer, realized through atomic layer deposition, on top of which we deposit the rhodamine layer, followed by a final capping layer of PMMA that provides stability of the emitters (see Figure \ref{fig:1}B). We modify the distance between the emitters and the graphene by increasingly depositing additional layers of TiO$_2$. In this way, we measure the lifetime of the emitters in the vicinity of a single graphene flake, while gradually increasing the graphene-emitter distance (i.e., the TiO$_2$ layer thickness). We choose experimental parameters such that processes other than non-radiative coupling to graphene are much less likely to occur. First, the spacer layer thickness ($5-20\,$nm) is much smaller than the emission wavelength of the rhodamine molecules (650\,nm) and is electrically isolating, so that charge transfer between the emitters and graphene is prevented \cite{Konstantatos2012}. Second, we use graphene with low intrinsic doping such that plasmons are not excited at this wavelength \cite{Koppens:2011p3370}. A more detailed description of the experimental setup and sample fabrication is presented in the Supporting Information.

\begin{figure}
\begin{center}
\includegraphics*[scale=0.8]{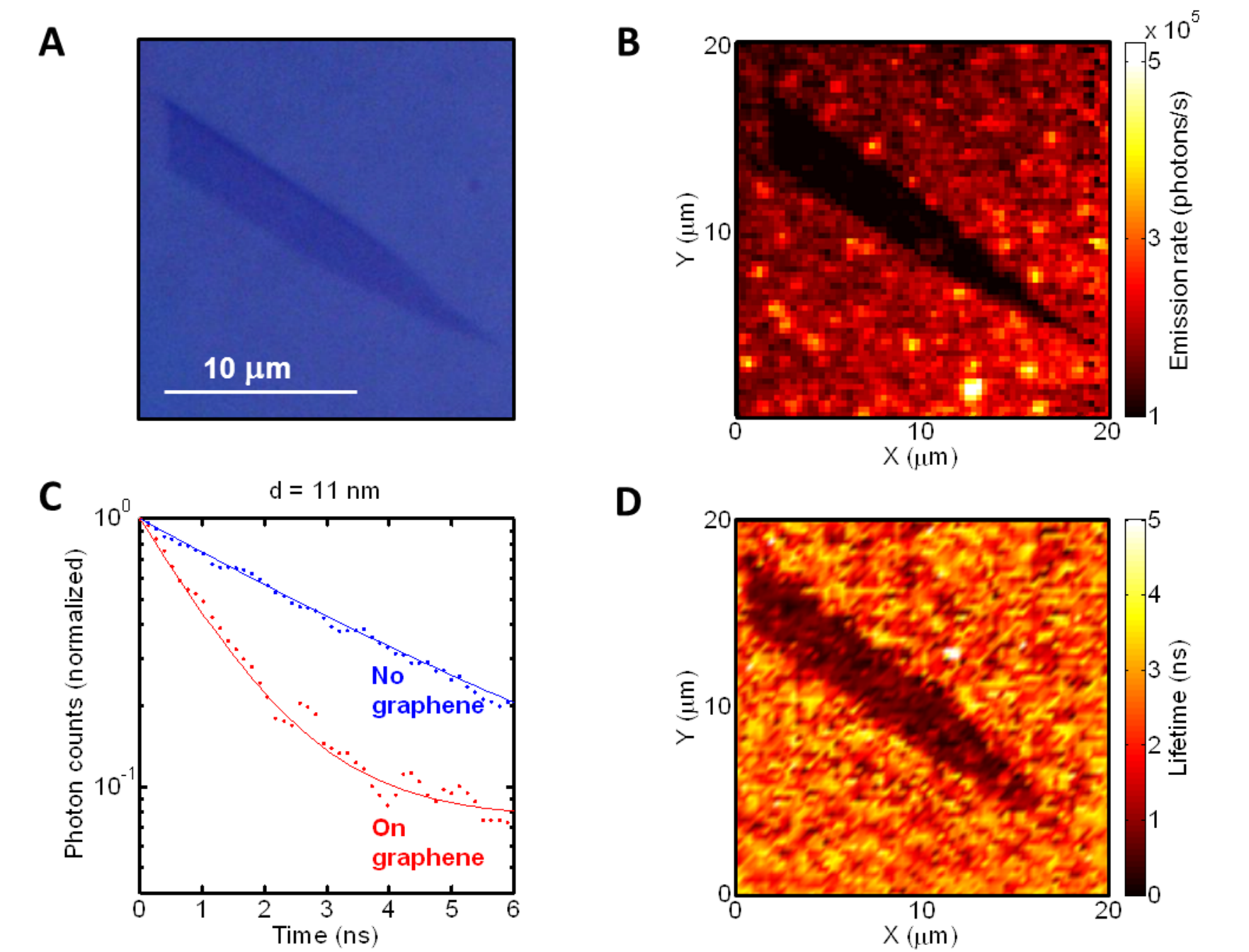}
\caption{(A) Optical microscope image of the single layer graphene flake (confirmed by Raman measurements) used in the experiments. (B) Fluorescence image of rhodamine molecules in the region depicted in panel (A). The distance separating between the emitters and the graphene flake is 11\,nm. Fluorescence quenching from the emitters on top of graphene is clearly observed. (C) Two emitter lifetime curves on a region above graphene (red) and not above graphene (blue). The solid lines are bi-exponential  fits to the data. (D) Lifetime image of the region depicted in panel (A), showing a clear reduction of the rhodamine lifetime in the region above the graphene flake.}
\label{fig:2}
\end{center}
\end{figure}

\section*{Results and discussion}

Figure \ref{fig:2}A shows an optical microscope image of the graphene flake used in the experiments, which can be correlated with Figure \ref{fig:2}B, containing a fluorescence image of the same area. Clear fluorescence quenching of the emitters that are located 11\,nm above the graphene sheet is observed, in agreement with earlier experimental works \cite{Treossi:2009gy,Kim:2010p3299,Huang2012,Chen:2010p1464}. This fluorescence quenching is due to the non-radiative energy transfer processes from the emitter to graphene.


Measurements of the lifetimes of the emitters respectively on top of graphene and outside graphene (above the substrate) are shown in Figure \ref{fig:2}C. As expected, the lifetime for emitters on graphene (defined as $\Gamma$)  is shorter compared with emitters on the substrate (defined as $\Gamma_{\rm s}$). Quantitative analysis (see Supporting Information) shows that there are two dominant contributors to the emission: the rhodamine emitters and a small background from the PMMA capping layer. Therefore, we fit the lifetime data with a double exponential decay $Ae^{-t/\tau_{\rm Rho}}+Be^{-t/\tau_{\rm PMMA}}$, where $t$ is the delay time between laser excitation and fluorescence, $\tau_{\rm Rho}$ and $\tau_{\rm PMMA}$ are the lifetimes of the rhodamine emitters and the PMMA protective layer, respectively, and A and B are the corresponding fluorescence intensities (in count rates). We find that, for all graphene-emitter distances, the emission from rhodamine is higher than that of the PMMA layer (A $>$ B) and that, due to the large thickness of the PMMA layer and therefore smaller coupling between graphene and PMMA, $\tau_{\rm PMMA}$ is independent of the distance, yielding a constant lifetime of $\tau_{\rm PMMA} =1.8$\,ns. The solid lines in Figure \ref{fig:2}C are fits according to this double exponential decay and yield rhodamine decay rates $\Gamma_{\rm{s}}=(3.5\,$ns$)^{-1}$ on the substrate and ${\Gamma_{\rm }}=(1.0\,$ns$)^{-1}$ on graphene, which give a decay rate ratio ${\Gamma_{\rm}}/{\Gamma_{\rm{s}}}$ of 3.5 for a graphene-emitter distance of 11\,nm. In Figure \ref{fig:2}D we show the extracted emitter lifetime $\tau_{\rm Rho}$ for the same region as the optical and fluorescence images of Figure \ref{fig:2}A,B. The good correlation between all three images confirms that it is the graphene flake that causes the enhancement of the decay rate of the emitters.

\begin{figure}
\begin{center}
\includegraphics*[scale=0.9]{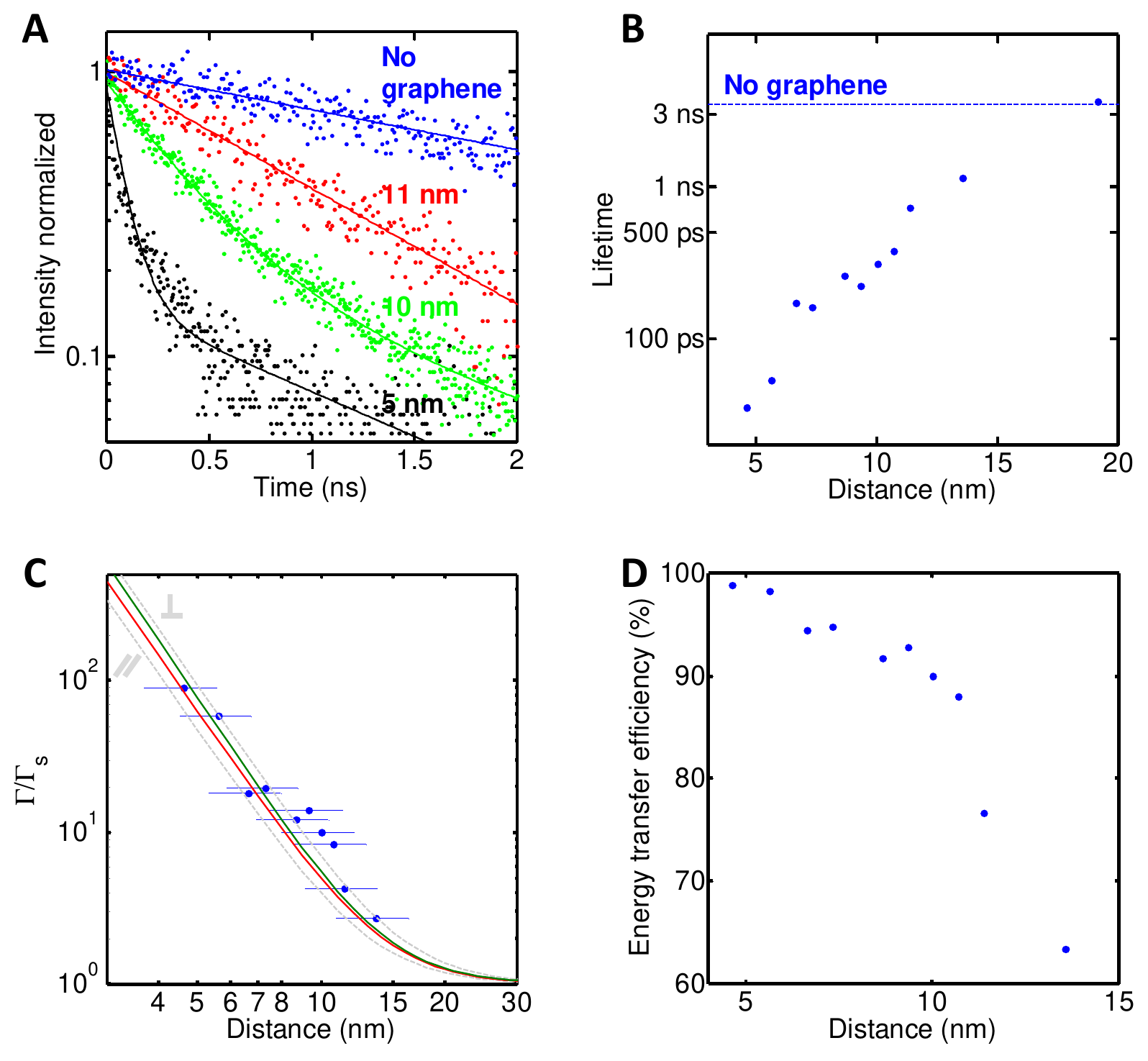}
\caption{Emitter lifetime reduction as a function of distance between the emitters and graphene. (A) Lifetime curves for emitters on top of graphene for distances of 5\,nm, 10\,nm, and 11\,nm, as well as for emitters outside the graphene. Solid lines are bi-exponential fits to the data. (B) Dependence of the rhodamine lifetime as a function of the distance separating them from the graphene. (C) Decay rate enhancement $\Gamma_{\rm }/ \Gamma_{\rm s}$ obtained from the lifetime measurements as a function of distance, yielding up to a factor 90. Dashed lines: analytical simulation for point-dipole emitters with their orientation parallel ($\Gamma_\parallel$) and perpendicular ($\Gamma_\perp$) to the graphene plane. The red solid line represents the weighted average over dipole orientations, $(1/3)\Gamma_{\perp}+(2/3)\Gamma_{\parallel}$. The green line represents the calculated $\Gamma_{\rm }/ \Gamma_{\rm s}$ (weighted average over dipole orientations) for the multilayer structure which includes the Si (0.5 mm), SiO$_2$(285 nm), TiO$_2$ (variable thickness), and PMMA (50 nm), with dielectric constants of 14.8, 2.12, 5.7, and 2.22 respectively \cite{Palik-Handbook}. (D) Energy transfer efficiency as a function of distance, calculated based on the lifetime measurements presented in (C). }
\label{fig:3}
\end{center}
\end{figure}

In order to produce a complete quantitative analysis and show the remarkably strong coupling between graphene and the emitters, we investigate how the ratio $\Gamma_{\rm }/{\Gamma_{\rm{s}}}$ depends on the emitter-graphene distance (see Figure \ref{fig:3}). In panel A, we present three lifetime measurements with a spacer layer of 5\,nm, 10\,nm, and 11\,nm, respectively, as well as the lifetime measured on a region without graphene, showing a clear reduction in lifetime as the emitters are placed closer to the flake. We repeated the measurement of the emitter lifetime outside the graphene flake region for all thicknesses of the spacer layer $d$ and found a lifetime of $3.5\pm0.3$\,ns that did not depend on the thickness of the TiO$_2$ spacer layer. The analysis indicates that the lifetime of the emitters increases with increasing distance until the emitters and graphene are separated by 20\,nm (Figure \ref{fig:3}B), where the measured lifetime contrast between emitters on graphene and outside graphene disappears. The corresponding ratio ${\Gamma_{\rm }}/{\Gamma_{\rm{s}}}$, presented in Figure \ref{fig:3}C, shows remarkably strong lifetime modifications reaching up to 90 for a distance of 5\,nm, which is an extraordinary result considering that this lifetime reduction is due to a single layer of atoms. We calculate the energy transfer efficiency via the relation $\eta=1-\Gamma_{\rm s}/\Gamma_{\rm }$, as is customary in F\"{o}rster-like processes. We find very efficient energy transfer, yielding more than 99\% at short distances, as depicted in Figure \ref{fig:3}D. We also note that the high efficiency of the energy transfer spans over a large distance range, with values $>85$\% at distances up to more than 10\,nm between the emitters and graphene.

\section*{Comparison with theory}

We compare our experimental results for the distance scaling of the energy transfer with a semi-classical model of an emitter (energy donor) coupled to a neighboring material (energy acceptor). The emitters are described as classical dipoles and the emission rate is worked out from the power that these dipoles radiate \cite{Blanco2004}. Details of the model are given in the Methods. For tutorial purposes we first consider an emitter placed in vacuum at a distance $d$ from graphene supported on a medium of permittivity $\epsilon$, while rigorous results for the more complicated multilayer structured of the experiment are given below. In the range of distances $d=1-15\,$nm, we find the decay rate $\Gamma_{\rm g}$ from emitter to graphene to be well described by (see Methods)
\begin{equation}
\Gamma_{\rm g}/\Gamma_0\approx 1+\frac{9\nu\alpha}{256\pi^3(\epsilon+1)^2}\left(\frac{\lambda_0}{d}\right)^4,
\label{eq1}
\end{equation}
where $\lambda_0$ is the free-space emission wavelength, $\Gamma_0$ is the rate in vacuum, and $\nu=1$ ($\nu=2$) when the emitter dipole is oriented parallel (perpendicular) to the graphene. Interestingly, the energy transfer rate does not depend on any material-dependent parameter related to the graphene \cite{GomezSantos:2011p1506}. The origin of this \textit{universal} scaling of the energy transfer between an optical emitter and graphene lies in the universal value for the optical conductivity $e^2/h$, which is attributed to the gapless and 2D character of the lossy graphene system, as discussed in the Methods section.

In Figure \ref{fig:3}C we compare this theoretical result with the experimentally obtained lifetime data. In order to yield a more realistic comparison, we have straightforwardly extended the theory to include the multilayer environment of our samples as well as retardation effects following the methods of Ref. 27.
We find that the calculated lifetime ratio $\Gamma /\Gamma_{\rm s}$ from this extended theory (green line in Figure 3c) is comparable to Eq. \ref{eq1} for $d>4$ nm  (red line in Figure 3c). We find that theory and experiment are in excellent agreement both qualitatively and quantitatively without any fitting parameters.

Both the measured data and Eq. \ref{eq1}  show the expected $d^{-4}$ dependence for the energy transfer rate to a two-dimensional material. It is well established that F\"{o}rster-like energy transfer has a typical scaling law $d^{-n}$, where $n$ is determined by the dimensionality of the system \cite{Barnes:1998p1486}. For two single-point coupled dipoles, the emission rate follows a $d^{-6}$ dependence. If one of the dipoles is replaced by a line of dipoles, integration reveals a $d^{-5}$ scaling. Analogously, for a two-dimensional array of dipoles, a $d^{-4}$ scaling is obtained, and finally F\"{o}rster processes scale with $d^{-3}$ when a single dipole interacts with a bulk of dipoles. Therefore our result is in agreement within the error with the expected scaling, given the two-dimensionality of the graphene sheet.

The observed energy transfer rate of an emitter close to graphene is very strong and reaches up to 90 times the decay rate in vacuum for a distance of 5\,nm. This can be attributed to two factors: the two-dimensional character of graphene and the relatively high value for its optical conductivity (strictly speaking the real part of its conductivity, which accounts for ohmic losses). The latter is in particular high for graphene due to the absence of a band gap which leads to a stronger coupling between electron-hole pairs and optical fields, compared with materials which have a bandgap. This high decay rate in front of an atomically thin material is counterintuitive, as one would expect that bulk, lossy materials should be more effective in absorbing from a nearby emitter. 

To put these results for graphene in context, we compare the decay rates of an emitter coupled to graphene and other materials which have a semi-infinite or (hypothetical) thin-film configuration. The model for the other materials starts from the same general form as for graphene (see Methods section). For a semi-infinite material with permittivity $\epsilon_m$, we have (with the emitter in vacuum)
\begin{eqnarray}
   \Gamma_{\rm 3D}/\Gamma_0 &\sim& 1+ \frac{3\nu}{128\pi^3} {\rm Im}\left\{\frac{\epsilon_m-1}{\epsilon_m+1}\right\} \left(\frac{\lambda_0}{d}\right)^3,
\end{eqnarray}
while for thin films, we find 

\begin{eqnarray}
   \Gamma_{\rm 2D}/\Gamma_0 &\sim& 1+\frac{9\nu}{512\pi^3} {\rm Im}(\epsilon_m-1/\epsilon_m)\,t \frac{\lambda_0^3}{d^4},
\end{eqnarray}
where $t$ and $\epsilon_m$ are respectively the thickness and permittivity of the film . These equations reveal the $d^{-3}$ and $d^{-4}$ scaling for bulk and 2D-materials.

\begin{figure}
\begin{center}
\includegraphics*[scale=0.4]{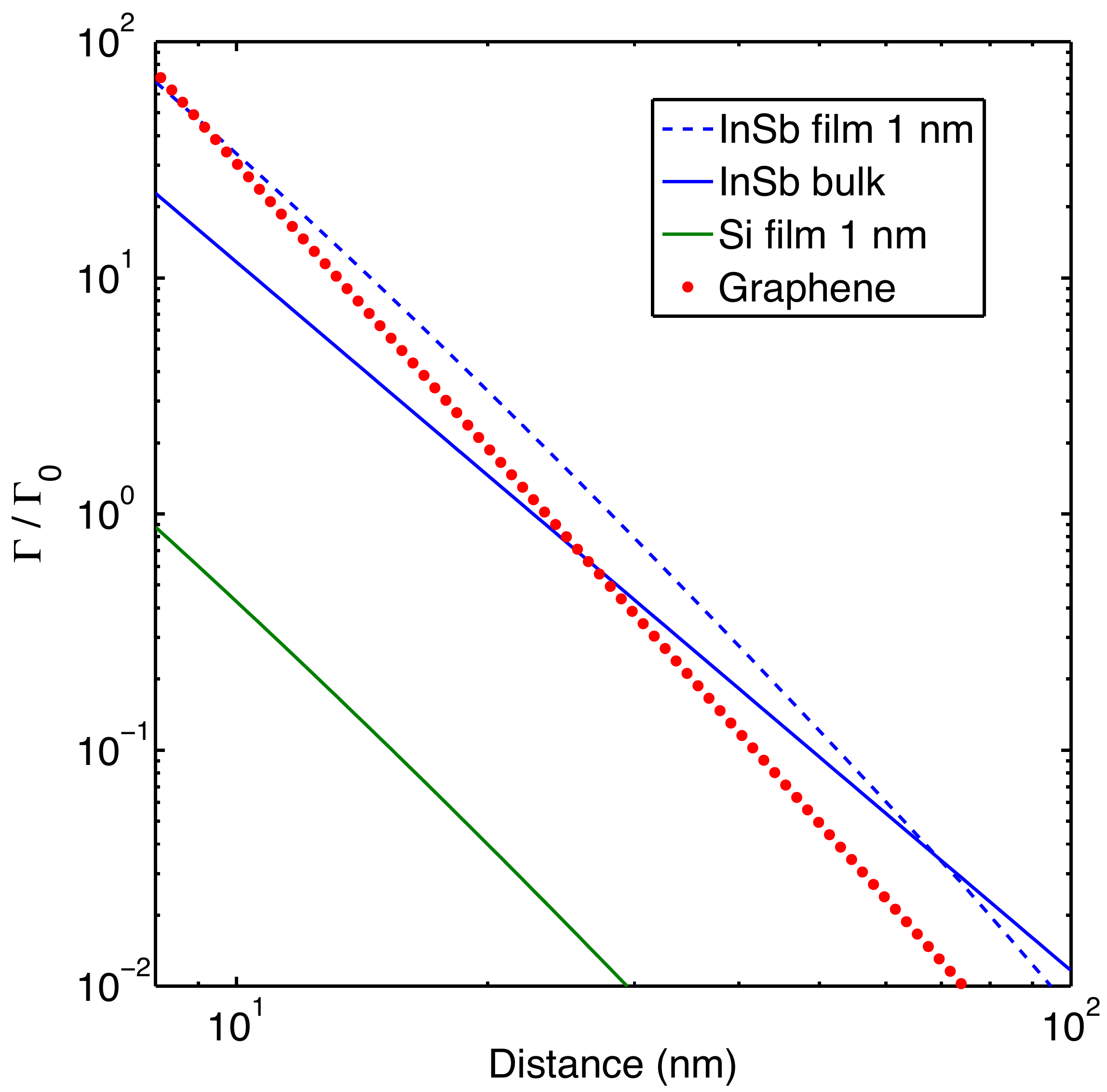}
\caption{{Decay rate enhancement versus distance for graphene, low-bandgap bulk InSb, and thin films of InSb and Si. The calculations are based on Equations (1) for graphene and Equation (2) for bulk materials, while for the thin films,  we use a similar  procedure as in the multilayer system \cite{Blanco2004}.}}
\label{fig:4}
\end{center}
\end{figure}

Tutorially, and n order to provide a more quantitative comparison, we calculate the decay rates of emitters close to both graphene, semi-infinite materials and thin films of 1\,nm. Figure \ref{fig:4} shows these decay rates for three different  materials: graphene, silicon, and InSb. The latter is a highly absorbing semiconductor because of its small band gap. These calculations provide us with two important insights. First, for smaller distances ($<10-50$\,nm, depending on the material), thin films lead to higher energy transfer rates than bulk materials. This is due to the crossover of the $d^{-4}$ scaling for 2D systems compared with the $d^{-3}$ scaling of bulk materials. Second, a (hypothetical) thin-film of a low band gap material shows similar energy transfer rates as graphene. This is because both graphene and low band gap semiconductors are very strong light absorbers and exhibit similar values for the real part of their optical conductivities. From the perspective of applications, our results show the interesting prospective of enhancing the energy transfer to a material by reducing its dimensionality, which can be readily achieved by using graphene. We note that for increasing distances, the optimal material thickness for highest energy transfer is a slightly thicker film than a monolayer, with a linear dependence of this optimal thickness on distance.
 
\section*{Conclusion}
\label{sec:Conclusion}

Our results show a strong enhancement of the energy transfer rate of emitters coupled to graphene as they come closer to the carbon layer, with a decay rate enhancement  factor of up to 90, and energy transfer efficiencies above 85\% for a distance up to 10\,nm. The experimental results are quantitatively consistent with a rigorous model based upon non-radiative energy transfer revealing a universal scaling of the decay rate of an emitter coupled to graphene, governed by the fine-structure constant $\alpha$ and the ratio $d/\lambda$.

The coupling between fluorescent materials and graphene is remarkably strong due to its two-dimensionality and gapless character. This can be exploited in numerous applications of varied nature. Noting that the dependence of the decay rate of the emitters versus distance is governed by material-independent fundamental constants and substrate optical properties, the graphene-emitter fluorescence can be used as an absolute nanoscale ruler \cite{GomezSantos:2011p1506}. Taking into account graphene's high mobility, it is foreseeable that combining highly absorbing emitters with graphene will result in highly efficient photo-detection and energy harvesting devices. In photonics, light can be shed on the dynamics of dark molecules (i.e., molecules with low quantum efficiency, whose ratio of intrinsic non-radiative to radiative decay rates is large). For example, if a dark molecules is coupled to graphene with an energy transfer rate that is larger than the intrinsic non-radiative decay rate, energy from the excited state of the dark molecule can be extracted, allowing for the study of excited state dynamics of dark molecules. Finally, the measured long range of the energy transfer can be used through DNA-length dependent fluorescence for biomolecule sensing applications with nanometer resolution \cite{Lu2009}.

\section*{Methods}

We use a semi-classical model of emitters coupled to a nearby material. The emitters are described as classical dipoles and the emission rate is worked out from the power that these dipoles radiate \cite{Blanco2004}. This produces results that are in agreement with a quantum-optics analysis \cite{Glauber1991}. For simplicity, we consider in this section the emission from an emitter placed in vacuum above an absorbing material. However, we have carried out a straightforward extension of this analysis to calculate the decay rate of an emitter in a multilayer system as that of the experiment (see Figure \ref{fig:1}A). The decay rate $\Gamma$ for an excited emitter in front of an acceptor material in the long-wavelength limit can be expressed as an integral over parallel wave vectors $k_\parallel$ as \cite{Blanco2004}
\begin{equation}
\Gamma/\Gamma_0=1+\frac{3\nu\lambda_0^3}{32\pi^3}\int_0^\infty k_\parallel^2dk_\parallel\,e^{-2k_\parallel d}{\rm Im}\{r_p\},
\label{eq:1}
\end{equation}
where $\Gamma_0$ is the rate in free space, $r_p$ is the $k_\parallel$-dependent Fresnel reflection coefficient for $p$-polarized light, $\lambda_0$ is the light wavelength, and $\nu=1$ ($\nu=2$) when the dipole is oriented parallel (perpendicular) to the surface. The distance dependence of the decay rate is related to the $k_\parallel$ dependence of $r_p$.

For graphene, we have ${\rm Im}\{r_p\}={\rm Im}\{\frac{-2}{\epsilon+1+4\pi\sigma ik_\parallel/\omega}\}$, where $\sigma$ is the graphene conductivity, $\omega$ is the photon frequency, and $\epsilon$ is the substrate permittivity. One can approximate \cite{Wunsch2006} $\sigma=e^2/4\hbar$, so that Eq. \ref{eq:1} admits a closed-form analytical solution $\Gamma/\Gamma_0=1+\nu C\,I(x)$, where $C=3(\epsilon+1)^2/2(\pi\alpha)^3$ is a constant and $x=[4(\epsilon+1)/\alpha](d/\lambda_0)$. Here, $I(x)=1/x^2+\mbox{Ci}(x)\cos(x)+\mbox{si}(x)\sin(x)$ is a function related to the $Ci$ and $si$ functions (tabulated in Ref. \cite{Abramowitz1965}) that can be approximated (within $<5$\% maximum relative error) as $I(x)\approx 1/(x^2+x^3/3+x^4/6)$. In the $3-15\,$nm range, this yields the result of Eq. \ref{eq1}, where we find a $\Gamma\sim d^{-4}$ dependence (similar to a conventional thin film). At small distances $d<3\alpha\lambda_0/[4(\epsilon+1)]\sim1\,$nm (below the range of our measurements), the first term in the denominator of $I(x)$ dominates, giving rise to a $\Gamma\sim d^{-2}$ dependence. At even smaller distances $d<v_F/\omega\sim0.3\,$nm, where the $k_\parallel$ dependence of $\sigma$ becomes relevant (nonlocal effects), one recovers a $\sim d^{-3}$ behavior typical of a semi-infinite medium.

For other materials, in the emission wavelength under consideration and at distances below $\sim20\,$nm, small $k_\parallel$ components dominate the integral of Eq. \ref{eq:1}, and we can approximate $r_p\approx g k_\parallel^n$ ($n=0$, 1, 2, or 3, depending on the dimensionality of the system), which leads to
\begin{equation}
\Gamma/\Gamma_0=1+\frac{3\nu\lambda_0^3}{32\pi^3}\frac{(n+2)!\;{\rm Im}\{g\}}{(2d)^{n+3}}.
\label{eq:2}
\end{equation}
Thus, the dominant distance-dependent term depends on the type of system under consideration. We examine the 3D and the 2D case. We find that in the 3D case ($n=0$) the Fresnel coefficient behaves as $r_p=g=(\epsilon-1)/(\epsilon+1)$, where $\epsilon_s$ is the permittivity of the substrate, leading to $\Gamma/\Gamma_0=1+C_{3D}(\lambda_0/d)^3$, where $C_{3D}=3\nu g/128\pi^3$. In contrast, for a thin film of permittivity $\epsilon$ and thickness $t\ll d$, we find $r_p=[(\epsilon^2-1)/2\epsilon] k_\parallel t$ (i.e., $n=1$), leading to $\Gamma/\Gamma_0=1+C_{2D}\,t\lambda_0^3/d^4$, where $C_{2D}=(9\nu/512\pi^3) {\rm Im}(\epsilon-1/\epsilon)$. The latter result has the same distance dependence as we find in graphene, which can then be ascribed to the 2D character of the carbon layer. In fact, we find from the above expression for $r_p$ in graphene that we can approximate ${\rm Im}\{r_p\}\approx(2\pi \alpha k_\parallel/k_0)/(\epsilon+1)^2$, where $k_0=\omega/c$ is the free-space light wave vector, under the assumption that $\pi\alpha k_\parallel/k_0\ll\epsilon+1$, thus recovering the same $n=0$ dependence as for an absorbing thin film.

The theory provided in Figure \ref{fig:3} is obtained without any fitting parameters. We find good agreement by representing the graphene through its DC conductivity $\sigma=e^2/4\hbar$.

{\bf Acknowledgment.} This work has been supported in part by the Fundacicio Cellex Barcelona, the ERC Starting grant no. 307806 (CarbonLight), the ERC Career integration grant 294056 (GRANOP), the Spanish MICINN (MAT2010-14885 and Consolider NanoLight.es), and the European Commission (FP7-ICT-2009-4-248909-LIMA and FP7-ICT-2009-4-248855-N4E). L.G. acknowledges  financial support from Marie-Curie International Fellowship COFUND and ICFOnest program. K.J.T. acknowledges financial support from NWO Rubicon grant.

\bibliographystyle{achemso}
\bibliography{./Lifetime-control_v25.bbl}

\end{document}